\pdfoutput=1 
\documentclass[sigconf]{acmart}
\usepackage{amsmath,amssymb,amsfonts,chemarrow,balance}
\usepackage{amsmath} 
\usepackage{graphicx}
\usepackage{subfigure}
\usepackage{mathenv}
\usepackage{color}
\usepackage{breqn}
 \usepackage{algpseudocode}
\newcommand{\BfPara}[1]{{\noindent\bf#1.}\xspace}
\usepackage{multirow}
\usepackage{float}
\usepackage{threeparttable}
\usepackage[inline]{enumitem}
\usepackage{wasysym}
\usepackage{etoolbox}

\usepackage{url}

\usepackage[T1]{fontenc}
\usepackage[utf8]{inputenc}
\usepackage{environ}
\usepackage{tikz}
\usetikzlibrary{arrows}
\usetikzlibrary{calc,matrix}
 
{\bfseries}{\itshape}
{\bfseries}{\itshape}
{\bfseries}{\itshape}
{\bfseries}{\itshape}
{\bfseries}{\itshape}
{\itshape}

\usepackage{xspace}
\usepackage{setspace}
\newcommand{\note}[1]{}

\newcommand{\etc}{{etc.}\xspace}
\newcommand{\eg}{{\em e.g.}\xspace}

\usepackage[inline]{enumitem}

\newcommand{\etal}{{\em et al.}\xspace}

\definecolor{xgreen}{rgb}{0.2,0.6,0.0}
\definecolor{xred}{rgb}{0.7,0.1,0.0}
\def\equationautorefname~#1\null{(#1)\null}

\usepackage{mathtools}  
\usepackage{amsmath}
\usepackage{amssymb}
\usepackage{tabulary}
\usepackage{amsmath}

\usepackage[linesnumbered,ruled,noend]{algorithm2e}
\usepackage{algpseudocode}
\usepackage{amsfonts}
\usepackage{booktabs}

\definecolor{light-gray}{gray}{0.95}
\definecolor{darkgray}{rgb}{0.4, 0.4, 0.4}
\definecolor{purple}{rgb}{0.65, 0.12, 0.82}
\definecolor{editorGray}{rgb}{0.95, 0.95, 0.95}
\definecolor{editorOcher}{rgb}{1, 0.5, 0} 
\definecolor{editorGreen}{rgb}{0, 0.5, 0} 
\definecolor{orange}{rgb}{1,0.45,0.13}      
\definecolor{olive}{rgb}{0.17,0.59,0.20}
\definecolor{brown}{rgb}{0.69,0.31,0.31}
\definecolor{purple}{rgb}{0.38,0.18,0.81}
\definecolor{lightblue}{rgb}{0.1,0.57,0.7}
\definecolor{lightred}{rgb}{1,0.4,0.5}
\usepackage{upquote}
\usepackage{listings}

\newcommand\JSONnumbervaluestyle{\color{blue}}
\newcommand\JSONstringvaluestyle{\color{red}}

\newif\ifcolonfoundonthisline

\makeatletter
\lstdefinestyle{json}
{  
    backgroundcolor=\color{light-gray},
    basicstyle=\footnotesize\ttfamily,
    showstringspaces=false,
    breaklines=true,
    frame=lines,
  showstringspaces    =false,
  keywords            = {false,true},
  commentstyle= \itshape\color{codegreen},
  alsoletter          =0123456789., \ifcolonfoundonthisline\JSONstringvaluestyle\fi,
  MoreSelectCharTable =%
    \lst@DefSaveDef{`:}\colon@json{\processColon@json},
  keywordstyle        = \ttfamily\bfseries,
}
\newcommand\processColon@json{%
  \colon@json%
  \ifnum\lst@mode=\lst@Pmode%
    \global\colonfoundonthislinetrue%
  \fi
}

\lst@AddToHook{Output}{%
  \ifcolonfoundonthisline%
    \ifnum\lst@mode=\lst@Pmode%
      \def\lst@thestyle{\JSONnumbervaluestyle}%
    \fi
  \fi
  \lsthk@DetectKeywords%
}

\begin{document}

\copyrightyear{2018}
\acmYear{2018}
\setcopyright{acmcopyright}
\acmConference[MobiQuitous '18]{EAI International Conference on Mobile and
Ubiquitous Systems: Computing, Networking and Services}{November 5--7, 2018}{New
York, NY, USA}
\acmBooktitle{EAI International Conference on Mobile and Ubiquitous Systems:
Computing, Networking and Services (MobiQuitous '18), November 5--7, 2018, New
York, NY, USA}
\acmPrice{15.00}
\acmDOI{10.1145/3286978.3286985}
\acmISBN{978-1-4503-6093-7/18/11}
\ccsdesc[500]{Security and privacy~Distributed systems security}

\title{Towards Blockchain-Driven, Secure and Transparent Audit Logs}

\settopmatter{authorsperrow=4}
\author{Ashar Ahmad}
\affiliation{%
  \institution{UCF}
}
\email{ashar@cs.ucf.edu}

\author{Muhammad Saad}
\affiliation{%
  \institution{UCF}
}
\email{saad.ucf@knights.ucf.edu}

\author{Mostafa Bassiouni}
\affiliation{%
  \institution{UCF}
}
\email{bassi@cs.ucf.edu }

\author{Aziz Mohaisen}
\affiliation{%
  \institution{UCF}
}
\email{mohaisen@cs.ucf.edu}

\begin{abstract}
Audit logs serve as a critical component in the enterprise business systems that are used for auditing, storing, and tracking changes made to the data. However, audit logs are vulnerable to a series of attacks, which enable adversaries to tamper data and corresponding audit logs. In this paper, we present {\em BlockAudit}: a scalable and tamper-proof system that leverages the design properties of audit logs and security guarantees of blockchains to enable secure and trustworthy audit logs. Towards that, we construct the design schema of {\em BlockAudit}, and outline its operational procedures. We implement our design on Hyperledger and evaluate its performance in terms of latency, network size, and payload size. Our results show that conventional audit logs can seamlessly transition into {\em BlockAudit} to achieve higher security, integrity, and fault tolerance.

\end{abstract}

\keywords{Blockchain, Audit Log, Hyperledger, Distributed Systems}

\maketitle

\section{Introduction}\label{sec:introduction}
Enterprise business systems and corporate organizations maintain audit logs to enable continuous monitoring and transparent auditing of system events~\cite{Wee99,OlivierS99}. Federal laws and regulations, including  Code of Federal Regulations of FDA, Health Insurance Portability and Accountability Act, \etc, require organizations to maintain audit logs for data auditing, compliance, and insurance~\cite{RingelsteinS09}.

Secure audit logs enable stakeholders to audit the state of systems, monitor users' activity, and ensure user accountability with respect to their role and performance. Due to such properties, audit logs are used by data-sensitive systems for logging activities on a terminal database systems. Often times, audit logs are also used to restore data to a prior state after encountering unwanted modifications. These modifications may result from attacks by malicious parties, software malfunctioning, or simply user negligence. 

Audit logs use conventional databases as their medium for record keeping. As such, they utilize a client-server model of communication and data exchange. Due to this client-server model, audit logs are vulnerable to a single point of trust, in which an adversary may access the database and/or the audit logs and manipulate critical audit information. As such, there is a need for secure, replicated, and tamper-proof audit logs that do not suffer from this shortcoming, and have effective defence capabilities to resist attacks. To that end, we envision that blockchain technology can naturally bridge the gap and nicely serve security requirements in audit log management, including ensuring provenance and transparency.

Recently, research on blockchains has gained significant momentum with applications spanning cryptocurrencies, health care, and IoT \cite{BonneauMCNKF15,GuoSZZ18,JesusCAR18}, among others . Blockchains enable secure, transparent, and immutable record keeping in distributed systems, without the need of a trusted intermediary. Applied to the audit log applications, blockchains can replicate audit logs over a set of peers, thereby providing them a consistent and tamper-proof view of the system. Furthermore, a malicious party aiming to hack the system will be required to change all the logs kept at each peer. This in turn, increases the cost and complexity of the attack, eventually increasing the overall defence capability of the audit log application.

Motivated by this, we present {\em BlockAudit}: an end-to-end solution that couples audit logs with blockchain systems and provides the design capabilities of audit logs as well as the security guarantees of blockchain systems. {\em BlockAudit} transforms the audit logs into a blockchain-compatible data structure. It then creates timestamped transactions from data within the audit logs and aggregates them in a block. It uses the energy efficient Byzantine Fault Tolerance (BFT) protocol to create consensus among the peers over the state of the blockchain. {\em BlockAudit} is application agnostic system that provides plug and play services to the audit log applications.

\BfPara{Contributions}
In summary, in this paper we make the following contributions. 
\begin{enumerate*}
    \item We highlight vulnerabilities in current audit log applications and present a threat model to captures those vulnerabilities.
    \item We present a system called {\em BlockAudit}, which defends against audit log attacks by leveraging the security properties of blockchains. 
    \item We provide the theoretical design of  {\em BlockAudit}, and use a real world eGovernment application, provided by Clearvillage Inc., to implement the system model on Hyperledger.
    \item Finally, we analyze the performance of our system using three evaluation parameters, namely the latency, network size, and payload size.
\end{enumerate*}

\BfPara{Organization}
The rest of the paper is organized as follows. 
In \textsection\ref{sec:background}, we provide the background and motivation of this work. In \textsection\ref{sec:thm}, we present the threat model, and in \textsection\ref{sec:cm} we propose the countermeasures. Results and evaluations are discussed in \textsection\ref{sec:eval}, followed by the related work and concluding remarks in \textsection\ref{sec:rw} and \textsection\ref{sec:conclusion}, respectively.

\begin{figure}[t]
\includegraphics[width=0.45\textwidth]{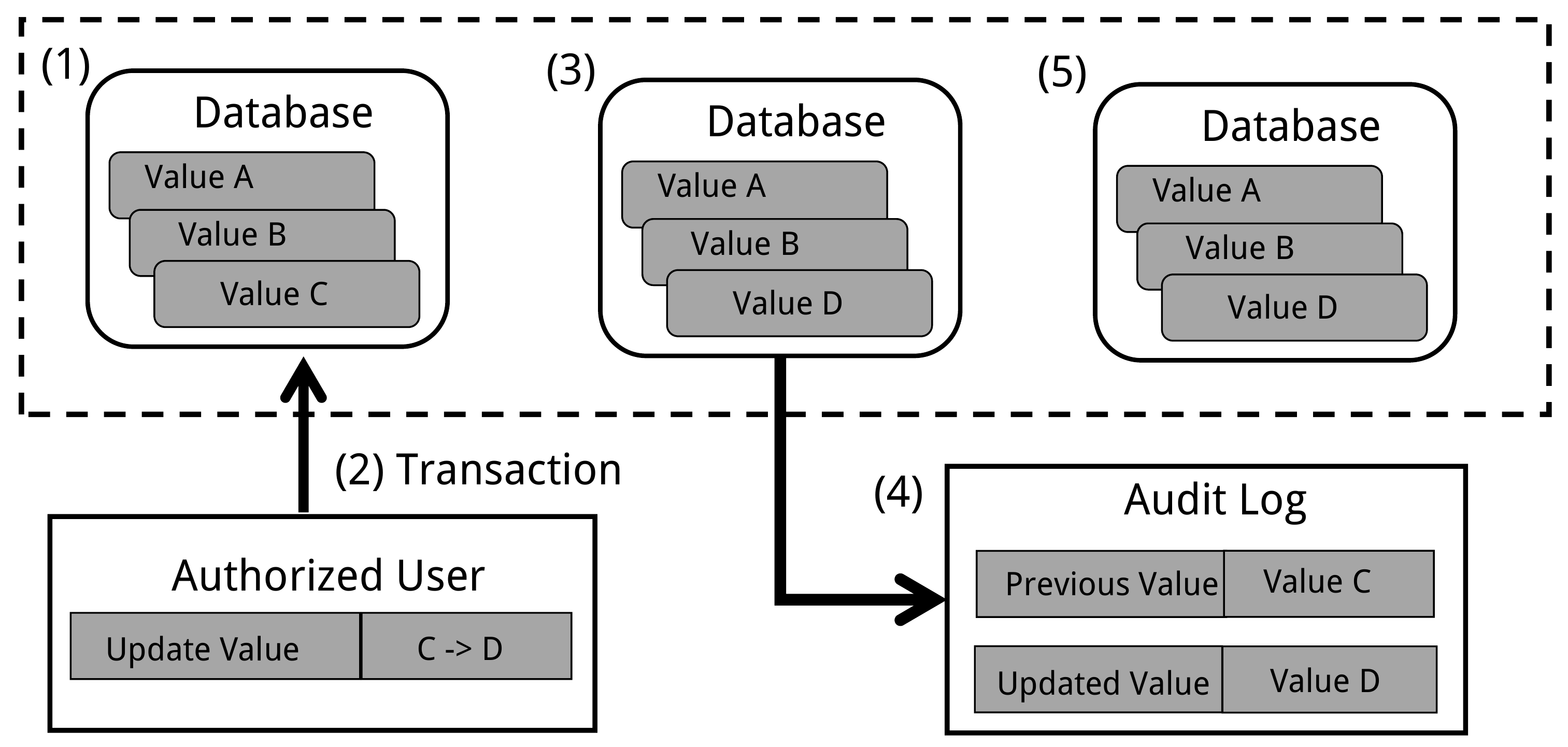}
\caption{Audit log generation in an OLTP system. We annotate each step with a number to show the sequence of progression. Notice that the user generates a transaction to change object value from C to D, and the change is then recorded in the audit log by the database.}
\label{fig:DesignOverview}
\vspace{-6mm}
\end{figure}

\section{Background and Motivation} \label{sec:background}
In this section, we provide the background  necessary to understand the operation of audit logs and blockchain.
\subsection{Audit Logs and Their Vulnerabilities} \label{sec:auditlogs}
Audit logs are an essential component in {\em online transaction processing} (OLTP) systems such as order entry, retail sales, and financial transaction systems \cite{VieiraM03,NikolaouMG97}. As such, the OLTP systems maintain audit logs to monitor users' activity and provide insights into sequential processing of transactions. Each processed payment in OLTP system creates a unique record in the audit log. The aggregate volume of transactions and the total payment made during a financial year can be verified by consulting the data recorded in the audit logs. Moreover, these audit logs can also be used to identify discrepancies, anomalies, and malicious activities in payments. Audit logs have to be searchable and easily accessible from the application, so that business users could look the chain of actions that resulted in the current state of the business object. In \autoref{fig:DesignOverview}, we provide an overview of OLTP system in which an audit log is generated once an authorized user commits a transaction to the database. The transaction makes a change in the value of an object and, as a result, the database records the change into an audit log. These changes can be later tracked and matched with the database for auditing and provenance assurance purposes.

\BfPara{Vulnerabilities in Audit Logs} 
Although audit logs are useful in tracking changes made to the database, and ensuring correctness in the system, they are  vulnerable to a series of attacks that may compromise the integrity of OLTP systems. An attacker can use multiple attack vectors to exploit known weaknesses in OLTP systems and corrupt the state of the database and the audit logs. Conventional schemes of protecting audit data include using an append-only device such as continuous feed printer or {\em Write Once Read Multiple} (WORM) optical devices. These systems work under a weak security assumption that the logging site cannot be compromised, which eventually keeps the integrity of the system intact. However, this weak notion of system security at the logging site is insufficient, and the attackers often have exploited vulnerabilities at logging site to tamper data in the audit logs \cite{LeeZX13,Margulies15}.

If an attacker gets hold of the credentials belonging to an authorized user, he can corrupt the database as well as the audit log. On the other hand, if the attacker compromises the database by breaching its defense, he can manipulate the database and prevent it from populating audit logs. Then, not only he will be able to corrupt the database, but also disable the auditing procedure by blocking the backward compatibility of audit logs with database.   

\subsection{Motivation} \label{sec:motivation}
Motivated by the use cases of audit logs and their security challenges, our aim is to come up with an end-to-end system that raises the security standards of existing OLTP systems while maintaining their operational consistency. We want to create a model that replicates the state of audit logs across multiple entities related to the OLTP application and maintain an append-only ledger. Moreover, the design should be robust against internal and external attacks on the database. If an attacker masquerades a legitimate party to gain the status of an authorized user, the user must be notified about the stolen credentials. If the attacker manages to infiltrate the database by exploiting an inherent vulnerability, then the distribution of audit logs across multiple peers must enable the detection of such an attack, and switch the system back to a stable state.

\section{Threat Model}\label{sec:thm}
In this paper, we assume an adversary capable of both physically accessing the trusted computing base (TCB) in the system and remotely penetrating the OLTP system by exploiting software bugs. We assume an OLTP system similar to a retail sale repository implementing the design logic of an application using secure communication protocols such as SSL/TLS. Moreover, the system has a database that keeps records of sales and maintains a remote audit log that keeps track of new changes made using transactions. In such a design, the attacker can exploit the system by launching two possible attacks: physical access and remote vulnerability attacks. 

\BfPara{The Physical Access Attack} 
In the physical access attack, the adversary will acquire credentials of a user that has the privilege of generating new transactions, and will use them to impersonate the legitimate user to corrupt the database. The attacker can manipulate data within the database as well as the audit log. Furthermore, the attacker will also be able to tamper the history maintained by the audit log in order to corrupt the auditing process. Therefore, in the physical access attack, we assume a strong adversary inside or outside the system who has access to the key system components. 

\BfPara{The Remote Vulnerability Attack}
In the remote vulnerability attack, the attacker may only exploit the default vulnerabilities in the OLTP applications such as software malfunctions, malware attacks, buffer overflow attacks \etc. In this attack, the adversary, although not as strong as the physical access attack, may still be able to contaminate the database and the audit log with wrong information. Despite these adversarial capabilities, we assume that the OLTP application is secure against conventional database and network attacks such as SQL injection and weak authentication.

\section{Countering Audit Log Attacks} \label{sec:cm}
To counter the aforementioned attacks, we propose a blockchain-based design called {\em Block Audit}, which integrates OLTP systems with blockchains guarantees to maximize the security of the system. 

Blockchain systems can be used to create consensus among peers in distributed systems \cite{CamenischDD17,saad2018poster}. In the following, we provide the design of {\em BlockAudit}, and show its applicability in OLTP systems.

\subsection{Block Audit} \label{sec:BA}
\subsubsection{Audit Log Application}\label{sec:do}
The first step in designing {\em BlockAudit} is the access to a large-scale audit log generation system that is currently being used by a an enterprise. For this purpose, we used the services provided by ClearVillage Inc. \cite{clearvillage}, which provides software for cities and counties, and maintains a searchable and persistent audit log to track changes in the database. 
Database is used for storing transactions, and searachable audit log is moved to Blockchain, this allows applications to continue have simple database queries, have sub-second response times, and benefit from the features of blockchain by using it to store the audit log. This can be done with minor modifications to existing applications.

\subsubsection{Application Architecture} \label{sec:dch}
Applications provided by ClearVillage use a multi-tier system architecture that comprises of web and mobile clients, a business logic layer, an object relational mapping (ORM), and a database. In the following, we describe the core functionality of each component along with the sequence of data exchange that eventually generates an audit log.  

\BfPara{Web Applications} Our web applications are built using asp.net, and the user accesses the application services through a web browser \eg Chrome. Additionally, native client support is provided for Android and iOS, built using their respective development frameworks. The web application and the web services are hosted on Microsoft’s Internet Information Services (IIS) web server.

\BfPara{Business Logic Layer} 
Business logic layer is an interface between the clients and the databases layer, responsible for implementing business rules. Among other functionalities, the business logic layer also manages data creation, data storage and changes to the data with the help of ORM. Upon receiving a request from the client, the webserver instantiates the relevant objects in the business logic layer, which uses the ORM to send the processed object to the client. The ORM writes changes to the objects in the relational database management system (RDBMS) tables.

\begin{figure}
\centering
\includegraphics[width=0.55\textwidth]{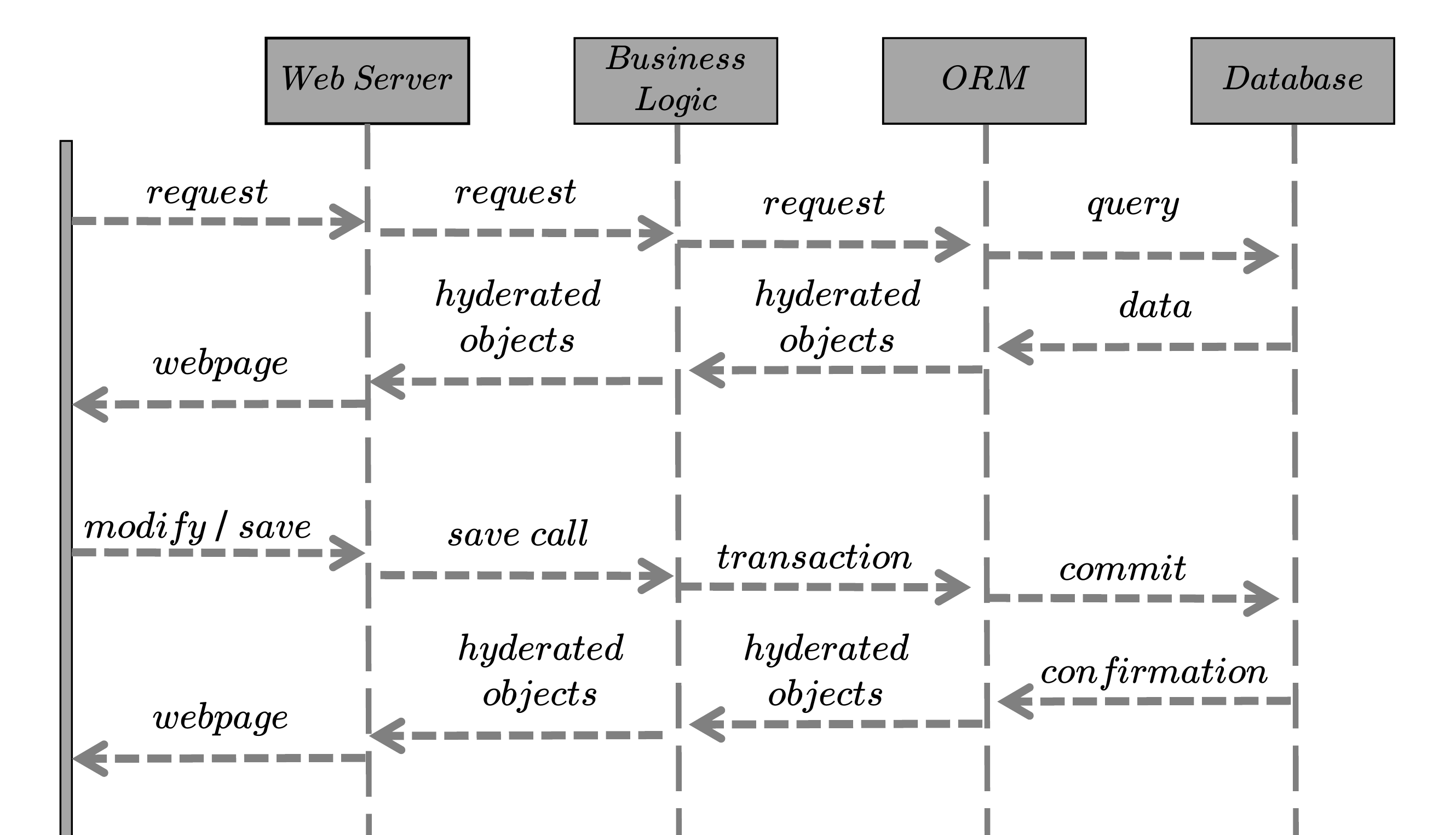}
\caption{The information flow between various components of the application. Notice that the transaction is generated at the business logic layer, and once the database commits to the transaction it is rendered on the webpage.}
\label{fig:ormmodel}
\end{figure}

\BfPara{ORM} The ORM in the application provides a mapping mechanism that allows querying of data from RDBMS using an object oriented paradigm. Modern web applications are well suited for this technique since they are multi-threaded and are rapidly evolving. ORM also reduces the code complexity and allows developers to focus on business logic instead of database interactions. This application uses {\em NHibernate}\cite{Nhibernate}: an ORM solution for Microsoft .NET platform. {\em NHibernate} is a framework used for mapping an object-oriented domain model to RDBMS, and it maps the .NET classes to database tables. It also maps Common Language Runtime (CLR) data types to SQL data types. The ORM inside the database layer creates an SQL statement to hydrated the object and passes it to the database. ORM also flushes the changes to the RDBMS, and commits a transaction. Interactions between the application and RDBMS are carried out using the ORM. In \autoref{fig:ormmodel}, we provide the information flow between various components of the application.   

\setlength{\textfloatsep}{1pt}
\begin{algorithm}[t]  
\SetAlgoLined\SetArgSty{}
\SetKwInOut{Input}{Input}

\SetKwProg{Fn}{Function}{}{end}
\Fn{OnPostInsert(PostInsertEvent e) }{
    \If{(e.Entity == AuditLog)}{\textbf{return;}} 

    \If{(e.Entity == AuditLogDetail)}{\textbf{return;}} 

    \If {e.HasAttribute(AuditableAttribute)}{\textbf{var new AuditLog(SessionId, AuditEventType.INSERT, EntityName, EntityId, UserId, Url);}}
      \For{$i = 0; i < e.Persister.PropertyNames.Length - 1 $ } {
         \If{(suppressedProp.Contains(propertyName))}{\textbf{continue;}} 
        
      auditLog.AddDetail(propertyName, oldValue, newValue);
      }
      \If{(auditLog.Details.Any()}{\textbf{SaveToBlockchain( auditLog);}} 
}

\caption{Creating of audit log entry for persisting new objects to database}
\label{algo:AuditCreate}
\end{algorithm}

\subsection{Generating Audit Logs}
In this section, we show how the application generates an audit log once the user commits a transaction. To Implement auditing  for this application, three events are used, namely {\em IPostInsertEventListener}, {\em IPostUpdateEventListener}, and {\em IPostDeleteEventListener}.

{\em IPostInsertEventListener} event is triggered once a transient entity is persisted for the first time. Each class that requires auditing is marked with {\em AuditableAttribute}, which is then used to create audit logs for classes containing this attribute. All mapped properties are then audited by default and a suppress audit attribute is added to suppress auditing of a target property. Usually, and by default, all properties are audited. However, in special cases where auditing is not required the suppress audit attribute is added to the property. In \autoref{algo:AuditCreate}, we show the process of generating the audit log when {\em IPostInsertEventListener} event is triggered.

When an audit entry is created, it contains a Session ID (transation ID), a class name, an event type (Insert, Update or Delete), audit ID, creation date, user ID, URL, and a collection of values for all properties. The collection of values consists of the old value before the update, and the new value resulting from the update. Moreover, during an update, the old and the new values are compared. Only if the two values are different from one another, that the change is committed to the audit log. In \autoref{algo:AuditUpdate}, we outline this procedure of updating the audit logs. Currently, these audit logs are saved inside an RDMBS using two tables, the AuditLog table and the AuditLogDetail table. Furthermore, the GUIDs in the audit log are used as primary keys.

Once the audit log is generated, the application provides a link to audit log page from the primary object. The link allows the end users to look at the history of the object, and track down any discrepancy caused by a system bug or malicious user activity.

\setlength{\textfloatsep}{1pt}
\begin{algorithm}[t]  
\SetAlgoLined\SetArgSty{}
\SetKwInOut{Input}{Input}

\SetKwProg{Fn}{Function}{}{end}
\Fn{OnPostUpdate(PostUpdateEvent e) }{
    \If{(e.Entity == AuditLog)}{\textbf{return;}} 

    \If{(e.Entity == AuditLogDetail)}{\textbf{return;}} 

    \If {e.HasAttribute(AuditableAttribute)}{\textbf{var new AuditLog(SessionId, AuditEventType.UPDATE, EntityName, EntityId, UserId, Url);}}
      \For{$i = 0; i < e.Persister.PropertyNames.Length - 1 $ } {
         \If{(suppressedProp.Contains(propertyName))}{\textbf{continue;}} 
         \If{(oldldValue <>newValue))}{\textbf{auditLog.AddDetail(propertyName, oldValue, newValue);}}  
      
      }
      \If{(auditLog.Details.Any()}{\textbf{SaveToBlockchain( auditLog);}} 
}

\caption{Creating of audit log entry for persisting existing objects to database}
\label{algo:AuditUpdate}
\end{algorithm}


\subsection{Blockchain Integration to Audit Logs} \label{sec:bcaudit}
In this section, we will show how the audit logs, obtained from our application, are integrated into  blockchain. So far in our design of {\em BlockAudit}, we have an application that stores audit logs upon receiving a transaction from a user. Now, we have to convert the audit log data into a blockchain-compatible format (blockchain transactions), and construct a distributed peer-to-peer network to replicate the state of the blockchain over multiple nodes.

\subsubsection{Creating Blockchain Network} \label{sec:cbn}
\begin{figure}[t]
\includegraphics[width=0.45\textwidth]{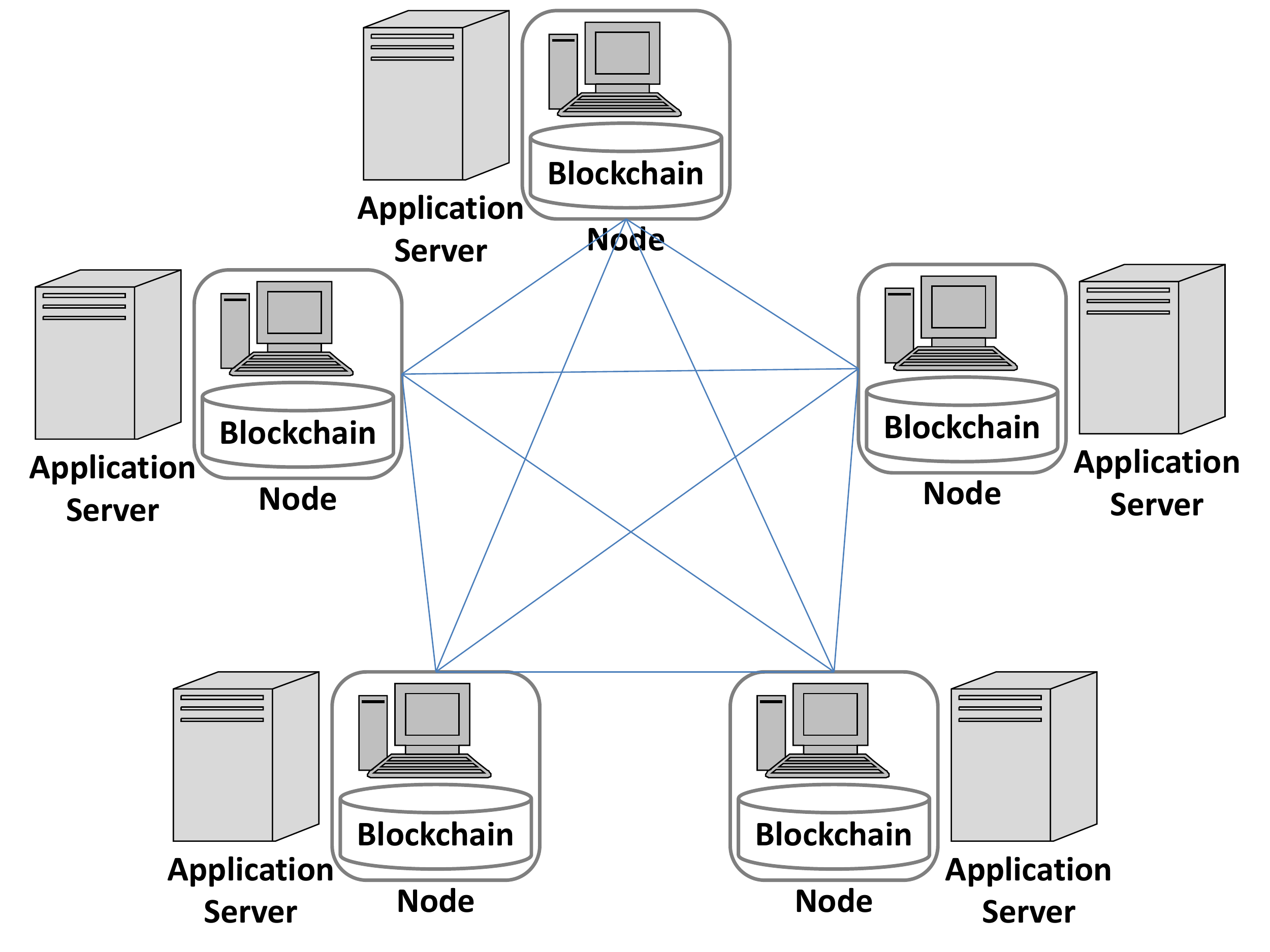}
\caption{ The network overview of nodes employing {\em BlockAudit}. Notice that each node maintains an interface that connects them to the auditlog application. They exchange transactions with one another during the application life-cycle. }
\vspace{-3mm}
\label{fig:server}
\end{figure}

The first step in our system deployment is setting up a distributed network of multiple hosts that process the blockchain transactions and create new blocks. In our {\em BlockAudit} design, the network consists of all peers that have the privilege of accessing the application and creating an audit log. Each node is required to keep a copy of the blockchain at their machines and maintain a persistent connection with the application server. Persistent connections are necessary to maintain an up-to-date view of the blockchain in order to process, validate, and forward transactions, as well as to avoid unwanted forks and partitioning attacks that may result from an outdated view of the blockchain. In \autoref{fig:server}, we illustrate the distributed architecture of nodes that reflect the peer-to-peer model of a blockchain application.

\subsubsection{Creating Blockchain Transactions} \label{sec:cbtx}
Once the network architecture is laid out, the next step is to create a blockchain-compatible transactions from the audit log data. For that, we convert the audit log data to a JavaScript Object Notation (JSON) format. We prefered JSON over other standard data storage formats such as XML, due to its data structure compactness and storage flexibility. To obtain a blockchain transaction, we first pass the audit log data to a function that serializes it to JSON,  and calls {\em createAudit} REST webservice to create the audit log transaction. Each JSON packet is then treated as a blockchain transaction, and as soon as a node in the network receives a transaction, it broadcasts it to the rest of the network. Nodes can connect to multiple peers to avoid the risk of delayed transactions due to malicious peer behavior or network latency. In \autoref{lst:json}, we show the data structure of the blockchain transaction that is obtained after serializing data from the audit log.

\begin{figure}[t]
\begin{lstlisting}[caption={Blockchain transaction generated after serializing data from the audit log. This transaction is exchanged among the peers during the application runtime},label={lst:json}, style=json]%[t]

{
   "ClassName":"SAGE.BL.InspSystem.PermitInspection",
   "CreatedDate":"\/Date(1532366360155-0400)\/",
   "EntityId":161031,
   "EventType":1,
   "Id":"9ceb8c2c-154a-49d5-9441-a92600db997b",
   "SessionId":"c66207c8-63be-4703-b858-cbfae98a988e",
   "Url":"\/SAGE\/Building\/Inspection\/InspectionReport.aspx?srcTp=309&srcId=17552018&InspectionTypeId=61663",
   "UserId":666,
   "Details":[  
      { 
         "Id":"fa268eaf-7993-48e3-ae6a-a92600db997b",
         "NewValue":"10",
         "OldValue":"9",
         "PropertyName":"DBVersion"
      },
      {  
         "Id":"ee2cdbc2-9c3a-4bc9-afba-a92600db997b",
         "NewValue":"only be available after 1:00 pm",
         "OldValue":"only be available after 2:00 pm",
         "PropertyName":"RequestComments"
      },
      {  
         "Id":"04b15535-7f8a-4899-8004-a92600db997b",
         "NewValue":"7\/23\/2018 1:19:20 PM",
         "OldValue":"7\/23\/2018 1:18:07 PM",
         "PropertyName":"LastUpdateDate"
      }
   ]
 }
\end{lstlisting}
\end{figure}

\subsubsection{Consensus Protocol} \label{sec:consptcl}
The next step in the design of {\em BlockAudit} is to employ a consensus scheme among the peers to formulate their agreement over the sequence of transactions and the state of the blockchain. There are various consensus algorithms used in blockchains, such as proof-of-work (PoW), proof-of-stake (PoS), proof-of-knowledge (PoK), byzantine fault tolerance (BFT), \etc \cite{SaadM18}. Each consensus scheme has its advantages and disadvantages. For example, PoW algorithms are highly scalable to a larger set of nodes, but have a high latency. In contrast, BFT protocols have a low latency, but suffer from scalability issues. For more details about consensus algorithms, we refer the reader to \cite{CromanDEGJKMSSS16}. 

For {\em BlockAudit}, we use Hyperledger, which employs variants of a BFT protocol to achieve consensus. Hyperledger is an enterprise permissioned distributed ledger that is developed by the Linux foundation and maintained by a strong steering committee. It has a growing development community that contributes towards the wide-scale adaptation of technology as a replacement of existing solutions. Our choice of Hyperledger is due to the design specifications of audit log applications, which require swift consensus while typically retaining a smaller sized network.  

\begin{figure}[t]
\includegraphics[width=0.45\textwidth]{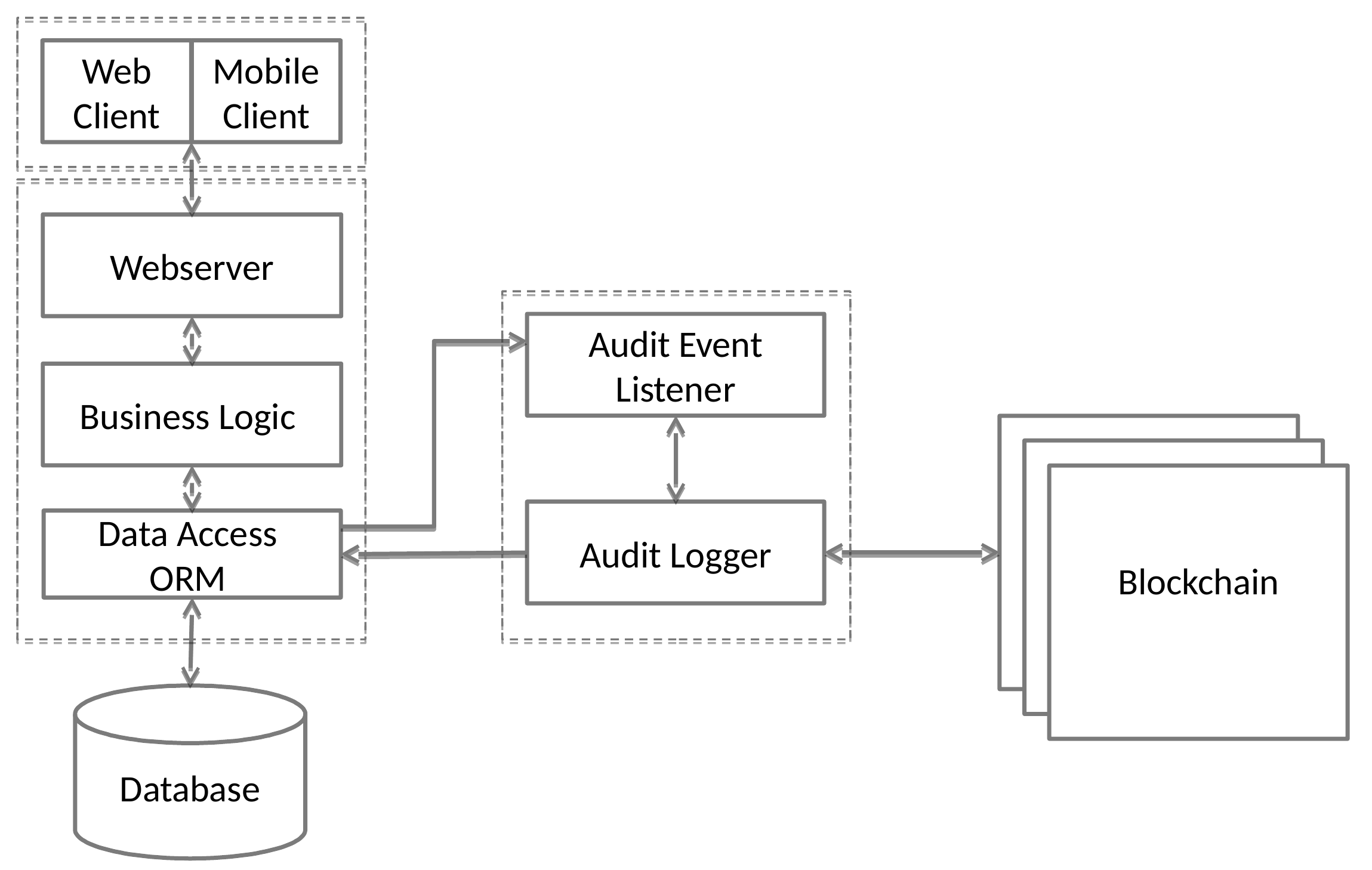}
\caption{Complete system architecture of {\em BlockAudit} after Hyperledger fabric is integrated to the JSON output. }
\vspace{-5mm}
\label{fig:archi}
\vspace{-3mm}
\end{figure}

Best suited to our current application, we decided to use Hyperledger Fabric \cite{AndroulakiBBCCC18}, which has a configurable and modular architecture that can be optimized for a diverse set of use cases. It also supports deployment of smart contracts atop blockchains in general-purpose programming languages; \eg Java, Go, and Node.js. Since Hyperledger Fabric is a permissioned blockchain, all network participants are known to one another, and there is a weak notion of anonymity in the system. The participants may not completely trust each other, as they might be competitors. However, by leveraging the capabilities of blockchains, trust can be augmented without the need of a third party. Additionally, Hyperledger fabric offers pluggable consensus, identity management, key management, and cryptographic libraries that can be customized based on the scope of the application. In \autoref{fig:archi}, we show the complete design of {\em BlockAudit}, once the Hyperledger fabric is integrated with the serialized JSON output of the business application.

\section{Experiment and Evaluation } \label{sec:eval}
We show the experiments carried out to evaluate the operation and performance of {\em BlockAudit}. We first extended the {\em nHibernate} ORM to generate a serialized JSON output in the form of a transactions as shown in \autoref{lst:json}. The transactions are broadcast to the network where a Hyperledger instance is configured at each node. For  experiments, we used HyperLedger Composer to set up the network and a Node.js wrapper to receive application's JSON transactions.

\begin{figure}[t]
\includegraphics[width=0.4\textwidth]{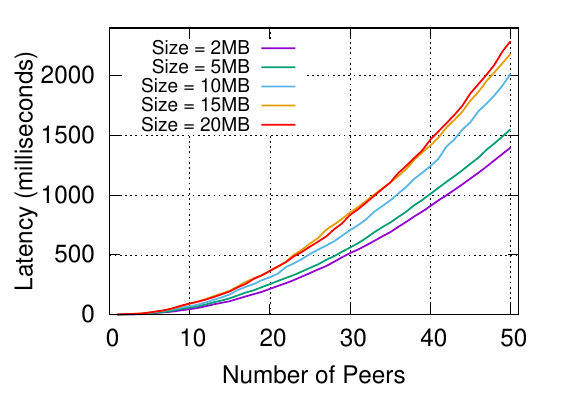}
\vspace{-3mm}
\caption{Performance of {\em BlockAudit}, as the payload size increases. Notice that as long as the network size remains within 30 peers, there is no significant change in the latency.}
\label{fig:performance}
\end{figure}

We evaluate the performance of our system by measuring the latency over the consensus achieved by the peers. We increase the payload size from 2MB to 20MB, and monitor the time taken for all the peers to approve the transactions. Let $t_{g}$ be the transaction generation time, and $t_{c}$ be the time at which it gets approval from all active peers and gets confirmed in the blockchain. In that case, the latency $l_{t}$ is calculated as the difference between $t_{c}$ and $t_{g}$ ($l_{t} = t_{c} - t_{c}$,  where $t_{c} > t_{g}$). We report our results in \autoref{fig:performance}, which shows that irrespective of the payload size, the latency margins remain negligible as long as the number of peers is less than 30. As the size of the network grows beyond 30 nodes, the latency factor increases considerably. Furthermore, we also notice, a sharp increase in latency when the payload size changes from 5--10MB and a negligible change in latency when the payload size changes from 15--20MB. These evaluation parameters can be used to define the block size and the network size, specific to the needs of the application. As part of our future work, we will use the parameters and the rate of incoming transactions to deduce optimum block size and the average block time for the audit log application.

\subsection{Discussion} \label{sec:discusion}
An essential component of our work is the defense against the attacks outlined in our threat model \textsection\ref{sec:thm}. In this section, we briefly discuss how {\em BlockAudit} is able to defend against the physical access attack and the remote vulnerability attack. In the physical access attack, if the attacker acquires credentials of a user, he can make changes to the application data using the application interface. In this case, his activity will be logged in {\em BlockAudit}. Since the log is kept in the blockchain by the user, the attacker will not be able to remove traces of his activity. Therefore, when the attacker's activity is exposed, auditors will be able to find out the effected records, and take corrective measures to restore data to the correct state. Moreover, if an attacker gets write access to the database, he might change data in different tables. Since the audit log is at the ORM, these changes will not be in the audit log, and will be detected.

In the case of a remote vulnerability attack in which the attacker exploits a bug or vulnerability in application, the audit log would show the effect of the changes or errors resulting from the attack. Additionally, the blockchain preserves the tamper-proof state of the audit log prior to the launch of the attack. As such, the auditors will be able to compare the audit log and the current data to spot the changes made during the attack. In the absence of the blockchain, if the attacker corrupts the prior state of the audit log, then there is no way the auditors can recover from it. However, with {\em BlockAudit}, not only the attacks are detected, but the system state is also recovered. Furthermore, for a successful attack in the presence of  {\em BlockAudit}, the attacker will need to corrupt the blockchain maintained by each node. Based on the design constructs and security guarantees of blockchains, this attack is infeasible.

\section{Related Work } \label{sec:rw}
In the following, we review the notable work done in the direction of secure audit logging mechanisms. We contrast their work with our approach to highlight our key contributions.

\BfPara{Audit Logs}
Schneier and Kelsey \cite{SchneierK99,SchneierK98} proposed a secure audit logging scheme capable of tamper detection even after compromise. However, their system requires the audit log entries to be generated prior to the attack. Moreover, their system does not provide an effective way to stop the attacker from deleting or appending audit records, which, in our case is easily spotted by {\em BlockAudit}. Snodgrass \etal \cite{SnodgrassYC04} proposed a trusted notary based tampering detection mechanism for RDBMS audit logs. In their scheme, a check field is stored within each tuple, and when a tuple is modified, RDBMS obtains a timestamp and computes a hash of the new data along with the timestamp. The hash values are then sent as a digital document to the notarization service which replies with a unique notary ID. The unique ID is stored in the tuple, and if attacker changes the data or timestamp, the ID received from the notary becomes inconsistent, which can be used for attack detection.

\BfPara{Blockchain and Audit Logs} Sutton and Samvi \cite{semwebSuttonS17} proposed a blockchain-based approach that stores the integrity proof digest to the Bitcoin blockchain. Castaldo \etal \cite{Castaldo2018} proposed a logging system to facilitate the exchange of electronic health data across multiple countries in Europe. Cucrull \etal \cite{CucurullP16} proposed a system that uses blockchains to enhances the security of the immutable logs. Log integrity proofs are published in the blockchain, and provide non-repudiation security properties resilient to log truncation and log regeneration.  In contrast, {\em BlockAudit} generates audit logs by extending the existing ORM (nHibernate), which is localized to ORM and other layers of business applicaiton are not effected. This makes it straightforward for existing application to user {\em BlockAudit}.

\section{Conclusion and Future Work} \label{sec:conclusion}
In this paper, we present a blockchain-based audit log system called {\em BlockAudit}, which leverages the security features of blockchain technology to create distributed, append-only, and tamper-proof audit logs. We highlight the security vulnerabilities in existing audit log applications, and propose a new design that extends ORM to create blockhain-compatible audit logs. For our experiment, we used an application provided by ClearVillage to generate transactions from audit logs, and record them in Hyperledger blockchain. By design, {\em BlockAudit} is agile, plug and play, and secure against internal and external attacks. In the future, we will extend the capabilities of {\em BlockAudit} by deploying it in a production environment, and explore various performance bottlenecks and optimization.  

\BfPara{Acknowledgement} This work is supported by Air Force Material Command award FA8750-16-0301 and Global Research Lab program of the National Research Foundation NRF-2016K1A1A2912757.
\vspace{-3mm}
\balance
\bibliographystyle{ACM-Reference-Format}
\bibliography{references,conf}

\end{document}